\documentclass[prl,twocolumn,showpacs,superscriptaddress]{revtex4}
\usepackage[dvips]{graphicx}
\usepackage{amsmath,amsfonts,amssymb}
\usepackage{dcolumn}
\usepackage{bm}
\usepackage{epsfig}
\def\beq{\begin{equation}}
\def\eeq{\end{equation}}

\def\subsc#1{{\mbox{\rm\scriptsize #1}}}

\def\Wcmcm{\mbox{\rm Wcm$^{-2}$}}

\def\N3d{N_\subsc{3D}}

\def\omegaMie{\omega_\mathrm{Mie}}

\def\omegalaser{\omega_\mathrm{l}}

\def\laserE{E_\mathrm{l}}

\def\laserEyz{E_{y,z}}

\def\Qb{Q_\mathrm{b}}
\def\ArN{\mathrm{Ar}_{N}}
\def\Arp{\mathrm{Ar}^{+}}

\def\omegaMie{\omega_\mathrm{Mie}}

\def\omegalaser{\omega_\mathrm{l}}

\def\Ionp{I_\mathrm{p}}
\def\chZ{\mathcal{Z}}

\def\vekt#1{\bm{#1}}
\def\vect#1{\vekt{#1}}
\def\vektr{\vekt{r}}

\def\vektR{\vekt{R}}

\def\vektE{\vekt{E}}

\def\vektEsc{\vektE_\mathrm{sc}}

\def\EhatYZ{E_{0y,z}}

\begin{document}
\title{
Harmonic Generation from Laser-Irradiated Clusters
}
\date{\today}
\author{M.\ Kundu}
\affiliation{Max-Planck-Institut f\"ur Kernphysik, Postfach 103980,
69029 Heidelberg, Germany}
\author{S.V.\ Popruzhenko}
\affiliation{Max-Planck-Institut f\"ur Kernphysik, Postfach 103980,
69029 Heidelberg, Germany}
\affiliation{Moscow State Engineering Physics Institute, Kashirskoe Shosse 31, 115409, Moscow, Russia}
\author{D.\ Bauer}
\affiliation{Max-Planck-Institut f\"ur Kernphysik, Postfach 103980,
69029 Heidelberg, Germany}

\date{\today}

\begin{abstract}
The harmonic emission from cluster nanoplasmas subject to short, intense  infrared laser pulses is analyzed by means of particle-in-cell simulations.
A pronounced resonant enhancement of the low-order harmonic yields is found when the Mie plasma frequency of the ionizing and expanding cluster resonates with the respective harmonic frequency.
We show that a strong, nonlinear resonant coupling of the cluster electrons with the laser field inhibits coherent  electron motion, suppressing the emitted radiation and restricting the spectrum to only low-order harmonics.
A pump-probe scheme is suggested to monitor the ionization dynamics of the expanding clusters.
\end{abstract}


\maketitle

During the last decade, the study of rare gas and metal clusters interacting with intense infrared, optical and ultraviolet laser pulses has emerged as a promising research field (see \cite{rost06} for a state of the art-review).
Generation of fast electrons and ions, production of high charge states, generation of x-rays and fusion in deuterium clusters have been observed in laser-cluster experiments with laser intensities up to $10^{20}$W/cm$^2$.
The hot and dense nonstationary, nanometer-sized plasmas far from equilibrium whose dynamics occur on the femtosecond  time scale---so-called {\it nanoplasmas}---are new physical objects with interesting properties very different from plasmas where finite-size effects can be neglected.
The probably most important feature of these nanoplasmas is the unsurpassed efficient energy transfer from the laser light to the charged particles \cite{ditmire97}.

While the energy absorption by cluster nanoplasmas has been widely studied both in experiments and theory, much less attention has been paid to the laser harmonic emission from such systems.
Meanwhile, it is known from the physics of intense laser-{\em atom} interaction that the effects of multiphoton absorption, leading to so-called above threshold ionization and high order harmonic generation, are intimately related and can be described as different channels of the highly nonlinear laser-atom coupling.
In intense fields, the laser-cluster coupling is also known to be highly nonlinear.
In fact, nonlinear resonance \cite{nl} has been shown (both in simulations and simple analytical models \cite{mulser05,kost05,antonsen05,kundu06}) to be of particular importance for the energy transfer from the laser pulse to the electrons of the nanoplasma and the subsequent outer ionization.
Most naturally the question arises whether  laser-driven clusters can be an efficient source of high-order harmonics as well.
Up to now, very few experimental results on harmonic generation (HG) from cluster targets have been published.
In Refs.\,\cite{ditmire96,ditmire97a,vozzi05} HG from rare-gas clusters irradiated by infrared pulses of moderate intensity ($\simeq 10^{13}-10^{14}$W/cm$^2$) was measured.
It was shown that under such conditions harmonics can be generated up to higher orders and with higher saturation intensities than in a gas jet.
However, the applied intensities were definitely not high enough to create a dense nanoplasma inside clusters so that the effects observed in Refs.\,\cite{ditmire96,ditmire97a,vozzi05} should be attributed to standard atomic HG modified by the fact that in clusters the atoms are disposed closer to each other
while the physical origin of HG remains the same as in a gas jet.
Only very recently, first experimental observations of the third harmonic (TH) generation from argon clusters subject to a strong laser field were reported in Ref.\,\cite{ditmire07}
where resonant enhancement of the TH yield, occurring when the Mie-frequency of the expanding cluster approaches three times the laser frequency, has been demonstrated using a pump-probe setup. The enhancement of the single-cluster response studied in theory before \cite{fom03,fomytski04,fom05} is, however, modified in the experiment by phase matching effects.
The latter point complicates experimental studies of nanoplasma radiation while in computations one can first examine the single-cluster response and may include propagation effects in a second step.

In this work, we concentrate on the single-cluster radiation in a short, intense laser pulse.
The main question we address here is whether the nonlinear dynamics of laser-driven nanoplasmas lead to a substantial harmonic emission and which conditions are required to optimize the signal.
We study HG by three-dimensional particle-in-cell (PIC) simulations for large $\ArN$ clusters (with the number of atoms $N\approx 10^4$--$10^5$ and radii $R_0\approx 6$--$10$\,nm) irradiated by linearly polarized, $n=8$ cycle sin$^2$-laser pulses with an electric field $\vect{\laserE}(t)=\vect{E}_0\sin^2(\omegalaser t/2n)\cos(\omegalaser t)$ and wavelength $\lambda = 800$~nm.
Previous PIC simulations of HG from laser-driven clusters \cite{antonsen05,fomytski04} were performed for two-dimensional (rod-like) clusters. 

In a first step the laser field ionizes all neutral atoms to $\Arp$ via over-the-barrier ionization once $|\vect{\laserE}(t)|\ge\Ionp^2(\chZ)\chZ^{-1}/4$ \cite{bethe}, where $\chZ$ is the charge number and $\Ionp(\chZ)$ is the respective ionization potential (atomic units are used).
Since the laser field alters the charge equilibrium a space charge field $\vektEsc(\vektr,t)$ is created.
Ions of higher charge states ($\chZ>1$) are produced if the condition $\vert \vect{\laserE}(t) + \vektEsc(\vektR_j,t) \vert\ge\Ionp^2(\chZ)\chZ^{-1}/4$ is satisfied at the individual ion locations $\vektR_j$.
Since a PIC electron has the same charge to mass ratio as a real electron
its equation of motion is $\ddot \vektr_i = - \vekt \laserE(t) - \vektEsc(\vektr_i,t)$.
Each PIC ion of  mass $M$ and charge $\chZ_j(t)$ moves according $M\ddot \vektR_j =\chZ_j(t)\left[\vekt \laserE(t) + \vektEsc(\vektR_j,t)\right]$.
The dipole approximation is applied, which is well justified for the laser wavelengths, cluster sizes and charge densities considered in this work.

\begin{figure}
\includegraphics[width=0.4\textwidth]{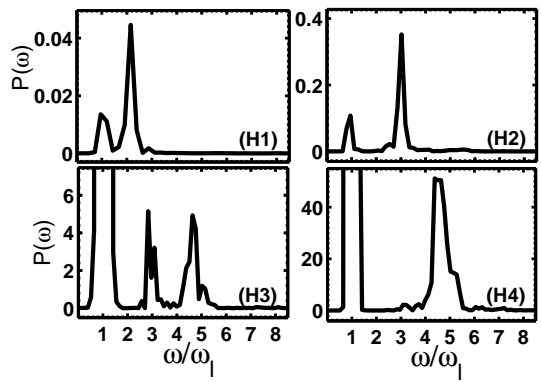}
\caption{Normalized harmonic power $P(\omega)=\vert A(\omega)/{N}\vert^2$ vs
harmonic order for an $\mathrm{Ar_{17256}}$ cluster at laser intensities
$2.5\times 10^{14}\Wcmcm$ (H1),
$2.5\times 10^{15}\Wcmcm$ (H2), and
$7.5\times 10^{17}\Wcmcm$ (H3).
The spectrum H4 is for an $\mathrm{Ar_{92096}}$ cluster at the intensity $7.5\times 10^{17}\Wcmcm$. 
\label{fig1}}
\end{figure}

Because of azimuthal symmetry the total dipole acceleration $A(t)$ is along the polarization direction  of the laser pulse ($x$-axis).
Its Fourier transformed amplitude $A(\omega)$ yields the dipole radiation power at the frequency $\omega$.
Figure~\ref{fig1} shows the normalized harmonic power $P(\omega) = \vert A(\omega)/{N}\vert^2$ vs
the harmonic order at various laser intensities.
The spectra H1, H2, and H3 are the results for an $\mathrm{Ar_{17256}}$ cluster ($R_0 = 6.2$~nm) while the spectrum H4 corresponds to an $\mathrm{Ar_{92096}}$ cluster ($R_0 = 10.9$~nm).
At the intensity $2.5\times 10^{14} \Wcmcm$ (H1) one observes a pronounced second harmonic.
Upon increasing the laser intensity the third harmonic power also increases and becomes comparable to the fundamental at higher laser intensities (H2).
Increasing the intensity further, the fifth harmonic appears in the spectrum (H3).
For the same intensity but a bigger cluster, the third harmonic is strongly suppressed as compared to the fifth harmonic (H4).

\begin{figure}
\includegraphics[width=0.4\textwidth]{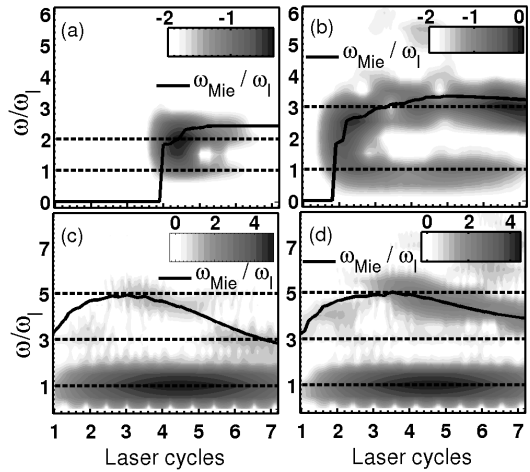}
\caption{Time-frequency diagrams corresponding to the spectra H1 -- H4 of Fig.\ref{fig1} with the normalized power $\log_{10}P$. 
\label{fig2}}
\end{figure}

The enhancements of particular harmonics are due  to the resonance between their frequencies and the Mie-frequency of the expanding nanoplasma \cite{fom03}.
To prove this we retrieve the temporal information of the radiation by a time-frequency (TF) analysis of the spectra H1-H4 in Fig.~\ref{fig1}.
Figure~\ref{fig2}a-d shows the respective TF diagrams.
The scaled Mie-frequency $\omegaMie(t)/\omegalaser$ vs time is included in the plots.
Note that a Mie-frequency can only be defined unambiguously as long as the ionic background remains homogeneously charged.
From the simulations we find that the charge homogeneity is well satisfied within the initial cluster radius $R_0$ while in the outer regions of the expanding cluster this is not the case.
We therefore define $\omegaMie(t) = \sqrt{\Qb(t)/R_0^3}$ with $\Qb(t)$ the total ionic charge inside the sphere of radius $R_0$ within which the cloud of well-bound electrons oscillates.

At the intensity $2.5\times 10^{14} \Wcmcm$ (Fig.~\ref{fig2}a, corresponding to H1 in Fig.\,\ref{fig1}) the laser field yields only $\mathrm{Ar}^{+}$ ions.
This first ionization by the laser field alone gives rise to an abrupt jump of $\omegaMie(t)/\omegalaser$ up to the value $\omegaMie/\omegalaser\approx 1.8$, followed by a slower increase above this value
due to ionization ignition \cite{rose97,bauer03} and finally ends with a plateau. The cluster expansion is so slow that $\omegaMie(t)/\omegalaser$ does not drop within the time interval plotted.
The second harmonic power peaks when $\omegaMie(t)/\omegalaser = 2$ is met.
This is because the electron dynamics contain oscillations both at the driving frequency and at the eigenfrequency, depending on the initial conditions and the form of the potential.
In long pulses such oscillations at the eigenfrequency are damped due to various relaxation processes and do not contribute to the measured harmonic signal.
Figure~\ref{fig2}b shows an enhanced third harmonic (H2 in Fig.\ref{fig1}) at the time when $\omegaMie(t)/\omegalaser\approx 3$.
The third harmonic power starts increasing again around the 6th cycle when $\omegaMie(t)$ passes through the same resonance due to the cluster expansion.
Figures~\ref{fig2}c,d show the TF spectrograms corresponding to spectra H3 and H4 (in Fig.\ref{fig1}) at the intensity $7.5\times 10^{17} \Wcmcm$.
Figure~\ref{fig2}c clearly shows enhanced third and fifth harmonic emission (at 3--4 and 6--7 cycles, respectively) when the scaled Mie-frequency approaches  the respective odd numbers.
For the bigger cluster Fig.\ref{fig2}d shows enhanced emission, preferentially following the Mie-frequency.
As in the case with the second harmonic shown in Fig.\ref{fig2}a this is a consequence of the undamped oscillations at the time-dependent eigenfrequency in the  expanding cluster potential. 
However, pronounced emission starts around the 4th cycle when $\omegaMie(t)$ meets $5\omegalaser$.
Despite the same laser intensity as in Fig.\,\ref{fig2}c $\omegaMie(t)$ does not reach $3\omegalaser$ during the slower expansion of the bigger cluster in Fig.\,\ref{fig2}d.
Note that the fulfillment of the resonance condition $\omegaMie(t) = m \omegalaser$ (with $m$ integer) is not sufficient for the emission of harmonics.
A necessary condition is the presence of some nonlinearity in the electron motion (i.e., anharmonicity in the effective potential).
This nonlinearity may either originate from  electrons jutting out of the core during their motion, sensing the Coulomb tail \cite{fom03}, or from the inhomogeneous charge distribution within the ion core \cite{fomytski04}.
\begin{figure}
\includegraphics[width=0.4\textwidth]{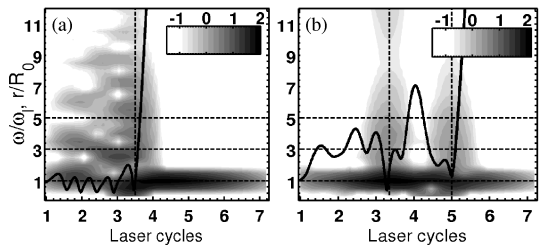}
\caption{TF spectrograms for two PIC electrons of the  Ar$_{17256}$ cluster irradiated by the laser intensity $5\times 10^{16}\Wcmcm$. The dimensionless electron distance from the cluster center $r(t)/R_0$ is indicated by the solid line.
    \label{fig3}}
\end{figure}

To clarify where the major contribution to the spectrum comes from we consider the radiation of individual PIC particles.
Figure~\ref{fig3} shows the TF analysis of the radiation from two PIC electrons. It is clearly seen that electrons radiate {\it harmonics} as long as they remain inside the cluster.
Leaving the cluster, they emit an intense flash with an almost continuous spectrum that extends up to significantly higher frequencies than present in the net harmonic spectrum shown in Fig.\ref{fig1}.
After liberation, the electrons emit only the fundamental frequency (linear Rayleigh scattering).
A liberated electron may rescatter, giving rise to a second flash in the TF spectrograms (Fig.\ref{fig3}b).
However, no indications of such intense flashes are visible in the net spectra shown in Figs.\,\ref{fig1} and \ref{fig2}.
The mechanism behind outer ionization (nonlinear resonance) allows to interpret these findings.
Individual electrons move in the selfconsistent field which could be subdivided into its slowly-varying and its oscillating component.
The slowly-varying field is induced by the quasistatic part of the space charge.
Ideally, it is a stationary potential well if one ignores a slow evolution of the charge distribution due to the inner and outer ionization and the cluster expansion.
The oscillating part of the field is a superposition of the incident laser field and the field induced by the oscillating electron cloud.
In the low-frequency limit $\omegaMie(t)\gg\omegalaser$ these two contributions are known to almost compensate each other so that the amplitude of the net oscillating field inside the cluster is small compared both to the applied laser field and the quasistatic space charge field.
If the electron energy in the quasistatic well is far from the resonance its trajectory remains weakly disturbed by the oscillating field.
This causes HG with rapidly decreasing yield as a function of the harmonic order so that even the 7th and 9th harmonic are barely present in the spectra.
As soon as the electron energy approaches the resonance, the same small perturbation results in a strong effect: the electron motion becomes strongly disturbed and stochastic.
A stochastic near-resonance behavior is a well-known property of nonlinear systems driven by  time-dependent forces \cite{liber}.
In the cluster case it leads to almost prompt and irreversible outer ionization.
Hence, passing through the resonance, an individual electron, upon leaving the cluster potential, emits radiation due to its strong acceleration, seen as a flash in the TF spectrograms (Fig.\,\ref{fig3}).
However, exactly because of the stochastic nature of nonlinear resonance  the electrons' trajectories are very sensitive to the initial conditions with which the nonlinear resonance is entered.
As a result, flashes from different electrons are incoherent (the corresponding amplitudes have nearly random phases), and, being added up in the total dipole acceleration, they disappear \cite{bandarage}. 
This shows that exactly the same mechanism behind efficient energy absorption by  and outer ionization from clusters, namely nonlinear resonance \cite{mulser05,kost05,antonsen05,kundu06}, restricts HG from them by breaking the coherent electron motion once it becomes strongly anharmonic.
Only well-bound electrons trapped inside the ionic core with energies far from the resonance contribute to the net, coherent radiation of the cluster.


The time-dependent enhancements of particular harmonics analyzed above may be used for the reconstruction of the maximum cluster charge density (i.e, maximum Mie-frequency). The TF spectrograms discussed above in Fig.\,\ref{fig2} were obtained from our numerical experiments while it is not possible to record them in a real ``single shot'' experiment. 
To that end a more realistic pump-probe setup (see, e.g., \cite{ditmire07,doeppner})  is envisaged with the following PIC simulations.
We  combine the $800$\,nm near-infrared laser pulse (M-pulse, pump) with two UV pulses (X-pulses, probe) $\laserEyz(t - t_d)=\EhatYZ\exp[-(\ln{2})(\omegalaser(t-t_d)/\pi)^2]
\sin[m\omegalaser(t-t_d)]$ of wavelength $\lambda_X = 800$\,nm$/m$ (with $ m = 3,5$), applied at different time delays $t_d$.
To distinguish the contributions of M and X-pulses to the dipole acceleration we apply them polarized along $y$ and $z$ axis, respectively.
The two X-pulses of different harmonic frequencies $m = 3, 5$ trigger independent Mie-oscillations of electrons in the respective directions.
The total power $P_{y,z}(\omega) = \vert A_{y,z}(\omega)/N\vert^2$ is recorded at various time delays.
Whenever $Q_b(t)$ is such that $\omegaMie(t) = m\omegalaser$ for a certain time delay  the corresponding detector is expected to measure the maximum dipole radiation.
Applying the two X-pulses simultaneously has the computational advantage of getting twice as much information in a single run than with just one probe pulse. 
Experimentally, a comb of low harmonics (e.g., from a gas jet) could be applied as a probe pulse.

\begin{figure}
\includegraphics[width=0.4\textwidth]{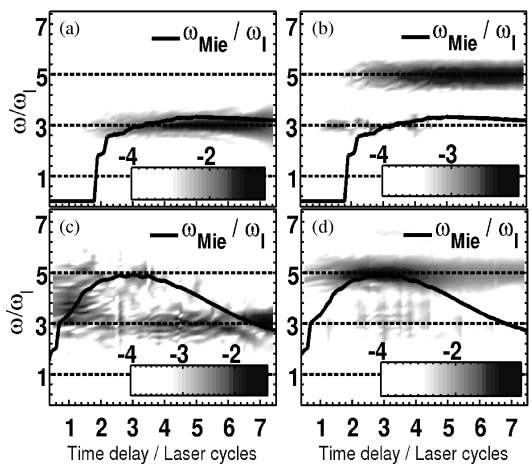}
\caption{
Spectrogram of the dipole power $\log_{10}P_{y,z}(\omega)$ along $y,z$ vs time delay $t_d$ and
frequency $\omega/\omegalaser$ with the pump-pulse intensities (a,b) $2.5\times 10^{15}\Wcmcm$ (cf.\ Fig.\ref{fig2}b), (c,d) $7.5\times 10^{17}\Wcmcm$ (cf.\ Fig.\ref{fig2}c) and the probe-pulse intensity $2.5\times 10^{12}\Wcmcm$ with $m = 3$ (a,c) and $m = 5$ (b,d).
\label{fig4}}
\end{figure}

Figures~\ref{fig4}a,b  show $\log_{10}P_{y,z}(\omega)$ vs $t_d$ and $\omega$ using X-pulses of frequencies $3\omegalaser$ and $5\omegalaser$, respectively. For $\omega_X = 3\omegalaser$ a resonance enhancement of the third harmonic power when $\omegaMie/\omegalaser$ crosses $\omega_X$ at $t_d>3.5$ cycles is clearly seen, showing that the information revealed in the TF spectrogram of Fig.\,\ref{fig2}b is accessible in the more realistic pump-probe set-up.
The enhancement is maximum when $\omegaMie/\omegalaser$ (i.e., the charge density $\rho$) is maximum around $t_d \approx 5$ laser cycles.
The frequency $\omega_X = 5\omegalaser$ of the X-pulse along $z$ in Fig.\ref{fig4}b, being off-resonant with $\omegaMie$, shows (an order of magnitude) less intense radiation.
Figure~\ref{fig4}a allows the conclusion that the maximum cluster charge density was at least in the vicinity corresponding to  $\omegaMie\approx 3\omegalaser$, i.e., $\max[\rho(t)] \approx 27\omegalaser^2/4\pi$ at that time.
Figures~\ref{fig4}c,d are the analogues of Figs.\ref{fig4}a,b but at an M-pulse intensity $7.5\times 10^{17}\Wcmcm$, corresponding to Fig.\,\ref{fig2}c.
The enhanced radiation is clearly seen at those times when $\omegaMie/\omegalaser$
meets the X-pulse (along $y$) frequency $\omega_X/\omegalaser = 3$.
However, the maximum of $\omegaMie/\omegalaser$  resonates with the other X-pulse (along $z$, Fig.\ref{fig4}d) frequency $\omega_X = 5\omegalaser$ at $t_d \approx 3$ cycles, and, indeed,  the maximum of the radiated power recorded in Fig.\ref{fig4}d is about an order of magnitude higher than in Fig.\,\ref{fig4}c. 

In conclusion, we investigated numerically harmonic emission from laser-driven cluster nanoplasmas.
The main contribution to the harmonic signal comes from electrons deeply bound inside the cluster potential.
Such electrons emit only low-order harmonics with considerable efficiency since they sense a weakly anharmonic potential only.
In contrast, electrons passing through the nonlinear resonance and leaving the cluster move along strongly disturbed trajectories. Their radiation, although intense and broad in wavelength, is incoherent due to the stochastic nature of the nonlinear resonance and therefore does not contribute to the net signal from the whole cluster.
The time-frequency analysis of the dipole acceleration shows enhanced third and  fifth harmonic emission when the Mie-frequency of the expanding cluster meets the respective harmonic frequencies, which is consistent with both previous theoretical studies \cite{fom03,fomytski04,fom05} and recent experiment \cite{ditmire07}.
A pump-probe experiment is proposed to measure the cluster charge density by detecting the dipole radiation at different time delays.
This method can be used to monitor the inner and outer ionization dynamics of clusters.  A control scheme to achieve the highest possible charge states by driving the cluster as resonantly as possible is currently investigated.

We acknowledge fruitful discussions with  W.\ Becker, A.\ Macchi, P.\ Mulser, and D.F.\ Zaretsky. This work was supported by the Deutsche Forschungsgemeinschaft.


\begin{thebibliography}{}
\bibitem{rost06} U. Saalmann, Ch. Siedschlag and J.M. Rost, J. Phys. B {\bf 39}, R39 (2006).

\bibitem{ditmire97} T. Ditmire, R.A. Smith, J.W.G. Tisch, and M.H.R. Hutchinson, \prl {\bf 78}, 3121 (1997).

\bibitem{nl} Here we call ``nonlinear resonance'' the first-order resonance between the laser field frequency $\omega_{\rm l}$ and the energy-dependent eigenfrequency of an individual electron trapped in an anharmonic, self-consistent potential.

\bibitem{mulser05} P. Mulser et al., \pra {\bf 71}, 063201 (2005);  \prl {\bf 95}, 103401 (2005).
    
\bibitem{kost05} I.Yu. Kolstyukov, JETP {\bf 100}, 903 (2005).

\bibitem{antonsen05} Th.M. Antonsen et al., Phys. Plasmas {\bf 12}, 056703 (2005).

\bibitem{kundu06} M.\ Kundu and D.\ Bauer, \prl {\bf 96}, 123401 (2006); \pra {\bf 74}, 063202 (2006).

\bibitem{ditmire96} T.\ D.\ Donnelly et al.,
	\prl {\bf 76}, 2472 (1996).

\bibitem{ditmire97a} J.W.G.\ Tisch et al.,
J.\ Phys.\ B {\bf 30}, 709 (1997).

\bibitem{vozzi05}  C.\ Vozzi et al.,
    Appl.\ Phys.\ Lett. {\bf 86}, 111121 (2005).

\bibitem{ditmire07} B. Shim et al., \prl {\bf 98}, 123902 (2007).

\bibitem{fom03} S.V.\ Fomichev  et al.,
J.\ Phys.\ B {\bf 36}, 3817 (2003).

\bibitem{fomytski04} M.V.\ Fomyts'kyi et al.,
Phys.\ Plasmas {\bf 11}, 3349 (2004).

\bibitem{fom05} S.V.\ Fomichev et al.,
J.\ Phys.\ B {\bf 37}, L175 (2004); \pra {\bf 71}, 13201 (2005).

\bibitem{bethe}  H.A.\ Bethe and E.E.\ Salpeter, {\em Quantum mechanics of
    one- and two-electron atoms} (Plenum Publishing Corporation, New York, 1977).


\bibitem{rose97} C. Rose-Petruck et al., \pra {\bf 55}, 1182 (1997).
\bibitem{bauer03}
	D.\ Bauer and A.\ Macchi, \pra {\bf 68}, 33201 (2003).



\bibitem{liber} A.J. Lichtenberg and M.A. Lieberman {\it Regular and Chaotic Dynamics}, (Springer, New York, 1982).

\bibitem{bandarage}  A similar behavior was observed in classical ensemble simulations of atomic HG  [Bandarage et al.,  \pra {\bf 46}, 390 (1992)]. 

\bibitem{doeppner} T.\ D\"oppner et al., \prl {\bf 94}, 013401 (2005); Eur.\ Phys. J.\ D {\bf 36}, 165 (2005). 

\end{thebibliography}
\end{document}